\documentclass[10pt, conference]{IEEEtran}

\IEEEoverridecommandlockouts                              
                                                          
\pdfoutput=1
\usepackage{cite}
\usepackage{amsmath,amssymb,amsfonts}
\usepackage{algorithmic}
\usepackage{graphicx}
\usepackage{textcomp}
\usepackage{amstext}
\usepackage{multirow}
\usepackage{threeparttable}
\usepackage{hhline}
\usepackage{subfigure}
\usepackage{url}
\usepackage{caption}
\usepackage{array}
\usepackage[linesnumbered,ruled]{algorithm2e}
\usepackage{xcolor}
\usepackage{float}
\usepackage{comment}
\usepackage{nicefrac}

\newcommand{\equref}[1]{Eq.~(\ref{#1})}
\newcommand{\figref}[1]{Fig.~\ref{#1}}
\newcommand{\tabref}[1]{TABLE~\ref{#1}}

\newcommand{\secref}[1]{Section~\ref{#1}}

\title{\LARGE \bf

Energy Harvesting Aware Multi-hop Routing Policy in Distributed IoT System Based on Multi-agent Reinforcement Learning \\\thanks{* Corresponding Author}

}

\author{\IEEEauthorblockN{Wen Zhang}
\IEEEauthorblockA{\textit{Department of Computer Science} \\
\textit{Texas A\&M University--Corpus Christi}\\
Corpus Christi, USA \\
wzhang3@islander.tamucc.edu}
\and
\IEEEauthorblockN{Tao Liu}
\IEEEauthorblockA{\textit{Department of Math and Computer Science} \\
\textit{Lawrence Technological University}\\
Southfield, USA \\
tliu3@ltu.edu}
\and
\IEEEauthorblockN{Mimi Xie}
\IEEEauthorblockA{\textit{Department of Computer Science} \\
\textit{University of Texas at San Antonio}\\
San Antonio, USA \\
mimi.xie@utsa.edu}
\and
\IEEEauthorblockN{Longzhuang Li}
\IEEEauthorblockA{\textit{Department of Computer Science} \\
\textit{Texas A\&M University--Corpus Christi}\\
Corpus Christi, USA \\
longzhuang.li@tamucc.edu}
\and
\IEEEauthorblockN{Dulal Kar}
\IEEEauthorblockA{\textit{Department of Computer Science} \\
\textit{Texas A\&M University--Corpus Christi}\\
Corpus Christi, USA \\
dulal.kar@tamucc.edu}
\and
\IEEEauthorblockN{Chen Pan*}
\IEEEauthorblockA{\textit{Department of Computer Science} \\
\textit{Texas A\&M University--Corpus Christi}\\
Corpus Christi, USA \\
chen.pan@tamucc.edu}
}

\begin{document}

\bstctlcite{IEEEexample:BSTcontrol}

\maketitle
\thispagestyle{empty}
\pagestyle{empty}

\begin{abstract}
Energy harvesting technologies offer a promising solution to sustainably power an ever-growing number of Internet of Things (IoT) devices. However, due to the weak and transient natures of energy harvesting, IoT devices have to work intermittently rendering conventional routing policies and energy allocation strategies impractical. 
To this end, this paper, for the very first time, developed a distributed multi-agent reinforcement algorithm known as global actor-critic policy (GAP) to address the problem of routing policy and energy allocation together for the energy harvesting powered IoT system. At the training stage, each IoT device is treated as an agent and one universal model is trained for all agents to save computing resources. At the inference stage, packet delivery rate can be maximized. The experimental results show that the proposed GAP algorithm achieves $\sim1.28\times$ and $\sim1.24\times$ data transmission rate than that of the Q-table and ESDSRAA algorithm, respectively.
\end{abstract}

\section{Introduction}\label{Introduction}

The emerging Internet of Things (IoT) is becoming increasingly popular. From wearable devices to environmental sensors, the growing number of IoT devices continue weaving a seamless web that benefits each corner of our daily life. There are more than 50 billion embedded IoT devices worldwide~\cite{IoT}. How to power those billions of distributed embedded IoT devices sustainably becomes as a pretty pressing issue. Out of all viable solutions, energy harvesting (EH) becomes one of the most preferable substitutes to the battery to addresses the cost, sustainability, and environmental concerns~\cite{pan2018enzyme,pan2019modeling}. However, due to the weak and transient nature of energy harvesting, IoT devices have to work intermittently, rendering conventional routing selection impractical. If a receiver experiences a power outage, the transmitter has to re-transmit all the data, resulting in unpleasant energy dissipation. Besides, routing loop avoidance is also challenging which needs frequent flooding~\cite{shin1993simple}. Yet, those methods require adequate and reliable energy supply, which is unattainable under EH environments.

Existing studies~\cite{rao2019renew, wang2020multi} have made efforts to improve the reliability and energy efficiency of routing selection by either finding the shortest routing path or minimizing the transmission time for the small-scale IoT systems powered by EH. However, the packet loss and energy depletion brought by the power intermittency are still yet to be solved. Since the data delivery is related to multiple tasks such as sensing and communication on embedded IoT devices, with a limited power budget, energy allocation for sensing will inevitably affect the energy allocation for communication making tasks on embedded IoT devices interdependent to each other. Such interdependence, however, offers us an opportunity to ameliorate the power intermittency problem. Specifically, by integrating energy allocation into routing selection, each embedded IoT device becomes capable of controlling the time and amount of data for communication with the proper energy allocation among runtime tasks.
Therefore, with the joint optimization of the routing selection and energy allocation, we can help maximize both the sink received data and the energy efficiency.

Nevertheless, the uncertainty of harvesting power creates a partially observable environment. In multi-hop networks, each node's current decisions influence the future. The current optimal energy allocation that maximizes the data transmission might result in insufficient energy for future data transmission. Moreover, the optimal decisions of a single IoT device also have an influence on its neighbors, which makes decisions of all IoT devices entangled. Therefore, whenever a single device is optimizing energy allocation, it also needs to consider whether this decision is long-term optimal or global optimal, which is crucial but has barely been explored. To this end, we are in a pressing need of joint-optimization of energy allocation and routing selection for each individual to address the complexity under this partially observable multi-hop environment. Of all viable solutions, Deep Reinforcement Learning (DRL)~\cite{mnih2015human} has been emerged as one of the best candidates. By naturally formulating a joint-optimization problem of energy allocation and routing selection as a partially observable Markov Decision Process (POMDP), DRL is able to make a sequence of decisions under uncertainty with an outstanding performance from a long term perspective.

Since decentralized networks are more commonly implemented in real-life IoT systems than well researched centralized networks, in this paper, for the very first time, a multi-agent DRL algorithm has been proposed to conduct joint-optimization for energy allocation and multi-hop routing to address the aforementioned concerns. In a nutshell, the major contributions of this paper are as follows:
\vspace{-4pt}
\begin{itemize}
  \item[1)] A comprehensive multi-hop data transmission models for energy-harvesting powered distributed IoT networks (as shown in~\figref{Fig:1}) has been developed.
  \item[2)] A global training distributed deploy multi-agent reinforcement learning (MARL) algorithm known as global actor-critic policy (GAP) has been proposed to address the aforementioned POMDP for the long-term optimality.
  \item[3)] A spatial reward mechanism is proposed for the entangled problem optimization of the multi-hop EH IoT system.
\end{itemize}

The remainder of this paper is organized as follows. \secref{rel} discusses related work. Then, \secref{sec:section2} builds an EH-powered distributed IoT system model. After that, \secref{sec:3} formulates the joint-optimization problem and proposes the corresponding DRL algorithm known as GAP. Finally, \secref{sec:4} conducts experiments to demonstrate the performance of GAP and \secref{sec:conclusion} concludes this work.

\section{Related Work}
\label{rel}
\textbf{Traditional Methods:} Nguyen et al.~\cite{nguyen2018distributed} proposed an EH routing algorithm for multi-hop heterogeneous IoT networks where the energy prediction model and the energy back-off parameters are integrated into the proposed routing algorithm.~\cite{banerjee2017joint} jointly optimizes the power allocation and routing selection for the EH Multi-hop Cognitive Radio Networks. However, these solutions are not designed for long-term optimization, resulting in throttled system performance.~\cite{lu2018energy} developed an algorithm, named ESDARAA, to explore multi-hop routing for EH IoT systems with energy-harvesting-aware geographic routing and different energy allocation strategies. However, it did not consider the uncertainty of the power source and the inter-dependency of each device.

\textbf{DRL-based Methods:} DRL shows outstanding performance on decision making in uncertain environments considering the long-term influence of its actions~\cite{tang2016reward,ortiz2017reinforcement,he2019routing}.~\cite{tang2016reward} proposed a multi-layer Markov fluid queue model to optimize the transmission by maximizing the reward for individual IoT devices rather than the multi-hop communication system. \cite{ortiz2017reinforcement} employed an algorithm to optimize the power allocation for two-hop EH communications. Q-table is created in~\cite{he2019routing} to find the optimal routing path for EH multi-hop cognitive radio network. 
However, these solutions only target on small-scale communication environments (i.e., at most $6$ nodes~\cite{he2019routing}), which is far away from the realistic IoT system that consists of a large number of interconnected nodes. Besides, few has considered synergizing energy allocation and routing for EH IoT systems. In this paper, GAP is compared with ESDSRAA~\cite{lu2018energy}, and Q-table, which represent energy-harvesting-aware routing, and energy-harvesting-aware routing with DRL, respectively.

\section{System Modeling}
\label{sec:section2}
In this section, a comprehensive network (\ref{sec:model_network}) and energy (\ref{sec:model_energy}), models will be given along with the problem formulation (\ref{sec:prob_fm}) for multi-hop EH IoT system. 

\subsection{Network Model}\label{sec:model_network}

\begin{figure}[b]\vspace{-10pt}
\centering
\vspace{-10pt}
\includegraphics[width=0.32\textwidth]{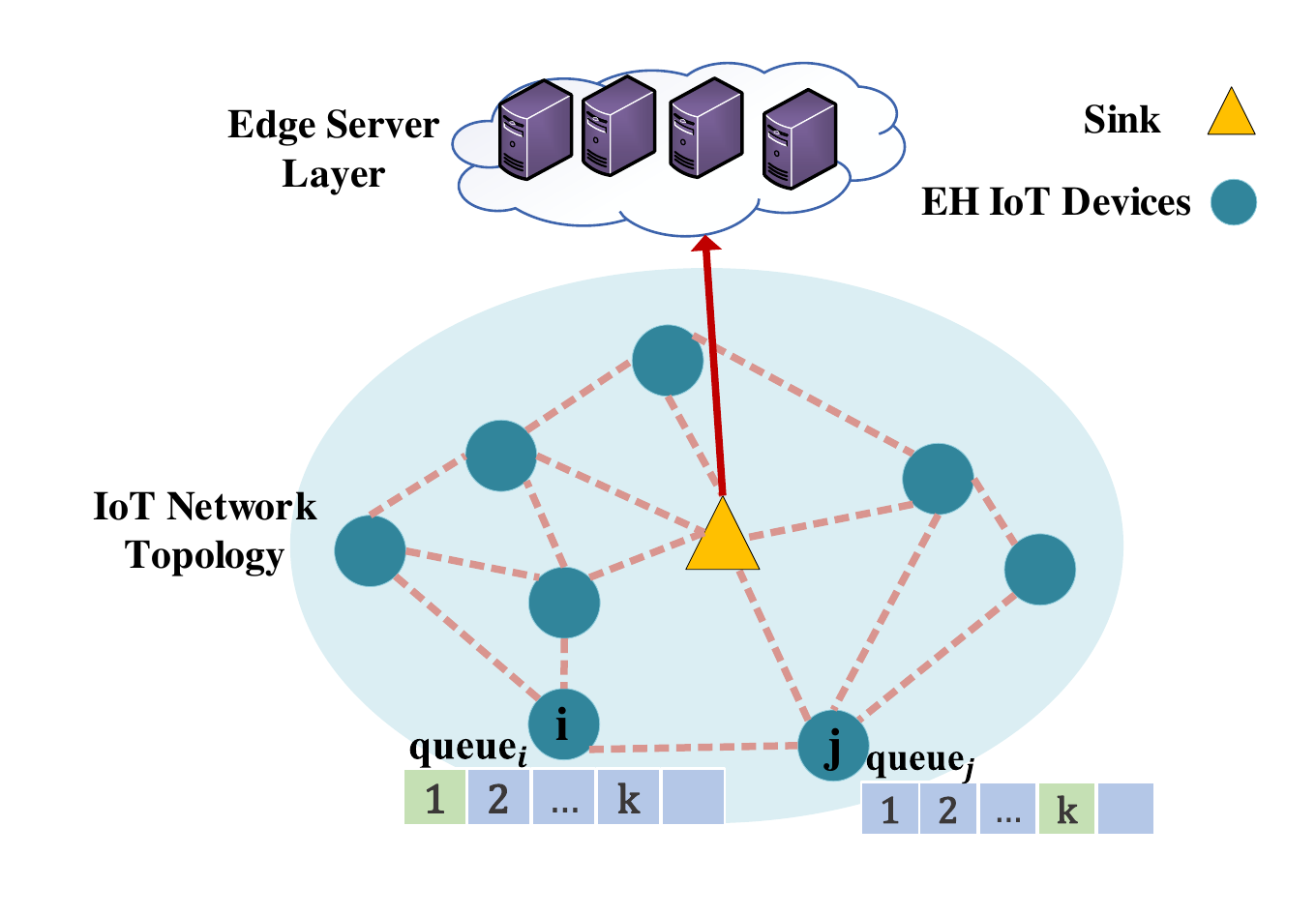}\vspace{-10pt}
\caption{Multi-hop EH IoT System.}
\label{Fig:1}\vspace{-5pt}
\end{figure}

In this subsection, the detailed network model is formulated~\cite{zhang2021sac}. Each IoT device has four runtime operations, including sensing, transmission, receiving, and energy harvesting. Due to the large network scale, Sink node ($\mathcal{O}$) is beyond the one-hop transmission range ($\zeta$) of most IoT devices. In this case, aside from sensing and transmitting its own data, each IoT device also conducts packet relay (receive and then transmit) for its one-hop neighbors, as shown in~\figref{Fig:1}.

\begin{figure}[b]\vspace{-10pt}
\centering
\vspace{-10pt}
\includegraphics[width=0.35\textwidth]{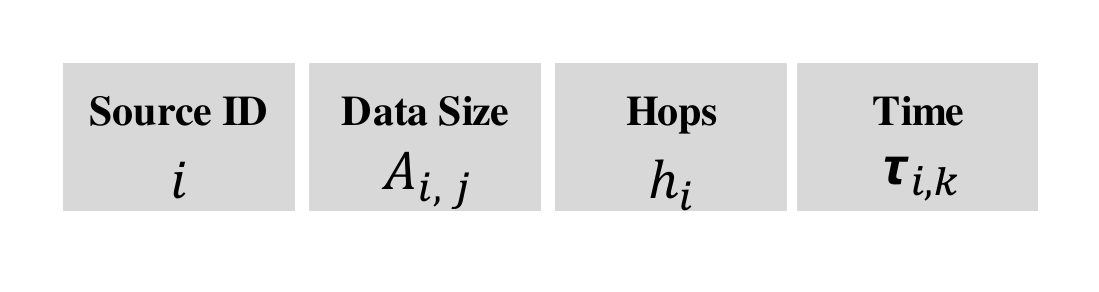}\vspace{-10pt}
\caption{Data Packet Structure.}
\label{Fig:packet}\vspace{-10pt}
\end{figure}

Assume that there are $M$ nodes in the network, where $i_{th}$ node is labeled as $i\in\mathcal{M}\{1,2,\cdots,M\}$. The neighbors of $i$ are stored in set $N_i$. 
While the node $i$ send data to $j$ we can retrieve the transmission rate $v_{i,j}$ in~\cite{he2019routing}.
Further, given $A_{i,j}$ as size of data transmitted from node $i$ to $j$. 
The transmission time $\tau_{i,j,trans}$ is calculated by $\tau_{i,j,trans} = \frac{A_{i,j}}{v_{i,j}}\quad i\in\mathcal{M},j\in N_i$.
There is a FIFO \textbf{queue} in each node to buffer the transmission data. The maximum queue size is $\Psi$ bits. Suppose that there are $k$ data packets in $\textbf{queue}_i$. The waiting time of the $k_{th}$ task is as follows:
\begin{equation}\small\vspace{-5pt}
  \tau_{k,wait} = \sum_{l=1}^{k}\tau_{i,j,sleep}^{l} + \sum_{l=1}^{k-1}\tau_{i,j,trans}^{l}\quad i\in\mathcal{M},j\in N_i,k\in\textbf{queue}_i
    \label{eq:3}
\end{equation}
Therefore, the total time costs of $k_{th}$ packet is $\tau_{i,k} = \tau_{k,wait} + \tau_{i,j,trans}^k\quad i\in\mathcal{M},j\in N_i, k\in\textbf{queue}_i$.

To avoid the routing loop and guarantee the data delivery before getting expired, the data packet structure is organized as~\figref{Fig:packet}, where $h_{i}$ is the packet's current number of hops when it arrives at node $i$ and $\tau_{i,k}$ is the time cost at node $i$. Given the maximum number of hops as $H$ and the expiration time as $T$. $h_i$ and $\tau_{i,k}$ can not exceed $H$ and $T$, respectively.

\subsection{Energy Model}\label{sec:model_energy}
Energy consumption mainly comes from four activities, including sensing, transmission, receiving, and sleep. Thus, the total energy cost $E_{i,con}$ for node $i$ can be given by $E_{i,con} =   \int P_{i,trans}\,dt + \int P_{i,recev}\,dt + \int P_{i,sleep}\,dt +  \int P_{i,sense}\,dt ,i\in \mathcal{M}$.
Further, the residual energy of $i$ is formulated as $E_{i,res} = E_{i,int} + E_{i,hav} - E_{i,con}, \quad i\in\mathcal{M}$,
where $E_{i,int}$ is the initial energy and $E_{i, hav}$ is the harvested energy that is equal to the power integral over time. With the maximum energy storage capacity as $E_{i,max}$, we have $E_{i,res} \leq E_{i,max}$.

\subsection{Problem Formulation}\label{sec:prob_fm}
This paper aims to maximize the received data on Sink within the expiration time. Therefore, the optimization problem can be described as follows: 
\begin{equation*}\vspace{-5pt}
\begin{aligned}\small
& \underset{i\in N_j, j = \mathcal{O}}{\text{maximize}}
& & \mathrm{\sum A_{i,j}} \\
& \text{subject to}
& & 0 \leq E_{i,res} \leq E_{i,max},h_{i}\leq H \quad\quad \forall i\in \mathcal{M} \\
& & & \sum_{l=1}^{k}A_{i,j}^l \leq \Psi \quad \forall i \in  \mathcal{M}, j \in N_i\\
& & & \tau_{i,k} \leq T \quad \forall i\in \mathcal{M},k\in \textbf{queue}_i
\end{aligned}
\end{equation*}

The constraints describes the EH IoT devices have a limited data storage and limited on-board power capacity.

\section{Gap Algorithm Design}
\label{sec:3}
In this section, we first briefly review policy-based DRL in section~\ref{sec:model_RL}. Based on the system models in section~\ref{sec:section2}, we then design a scalable and distributed multi-agent DRL algorithm, known as GAP, to optimize the energy efficiency of the EH IoT system for communication, which is shown in subsection~\ref{sec:design_IA3C}. After that, the detailed configuration of GAP will be discussed in subsection~\ref{sec:ia3cset}.

\subsection{Reinforcement Learning}\label{sec:model_RL}
Reinforcement learning formulates the targeting environment as a Markov Decision Process (MDP), where the agent captures the environment state ($s_t$) at time step ($t$), then follows a policy ($\pi$) to take action ($a_t$). Based on the result of environment transition, an immediate reward ($r_{t+1}$) and the observable environment state ($s_{t+1}$) will be given to the agent at $t+1$. The long-term maximized reward ($R_t^{\pi}$) can be found at $t$. It is formulated as $R_t^{\pi} = \sum_{\tau=t}^{\infty}\gamma^{\tau-t}r_{\tau+1}
$,
where the $\gamma \in (0,1]$ is a discounted factor with which the expected future reward can be formulated as a Q-function $ Q(s,a)^\pi = \mathbb{E}[R_t^\pi|s_t=s,a_t=a]$.
In DRL, the policy is defined as $\pi_\theta(a|s)$, where with the input environment state $s$, the parameter model $\theta$ will output the action $a$ that has the maximum expected future reward. Specifically, by integrating deep neural networks (DNN), $\theta$ is represented by the DNN inference model. Therefore, our goal, after training, is to obtain the optimal inference model $\theta^*$ with the corresponding optimal action  $a=argmax Q(s,a)^\pi_\theta$.

To improve the training performance, actor-critic method is adopted. First, we use a new neural network model $\Omega$ (critic) to estimate the Q-value $\mathbb{E}[R_t^{\pi}|s_t = s] $ conditioned on the state $s$ as V-function $V_{\Omega,s_t}$. The loss function to update $\Omega$ is~\eqref{eq:L_func_omega}.
\begin{equation}\small\label{eq:L_func_omega}
   \mathcal{L}(\Omega) = \nicefrac{1}{2B}\sum (R_t - V_{\Omega,s_t})^2 
\end{equation}
Then, the advantage action function, $Ad(s_t, a_t) = Q(s,a)^\pi - V^\pi(s_t)$, is adopted that indicate the advantage degree of action $a$. The advantage value at time step $t$ is denoted as $Ad_t = R_t-V_{\Omega, s_t}$. For the policy neural network $\theta$, we aims to optimize the probability so that the action with higher Q-value will has a higher probability to be selected. Hence, the loss function for $\theta$ can be formulated as in~\equref{eq:L_func_theta}. \begin{equation}\label{eq:L_func_theta}\small
   \mathcal{L}(\theta) = -\nicefrac{1}{B}\sum log \pi_{\theta}(a_t|s_t)Ad_t 
\end{equation}

\subsection{GAP for EH IoT System}\label{sec:design_IA3C}
\begin{figure}[t]\vspace{-20pt}
\centering
\includegraphics[width=0.52\textwidth]{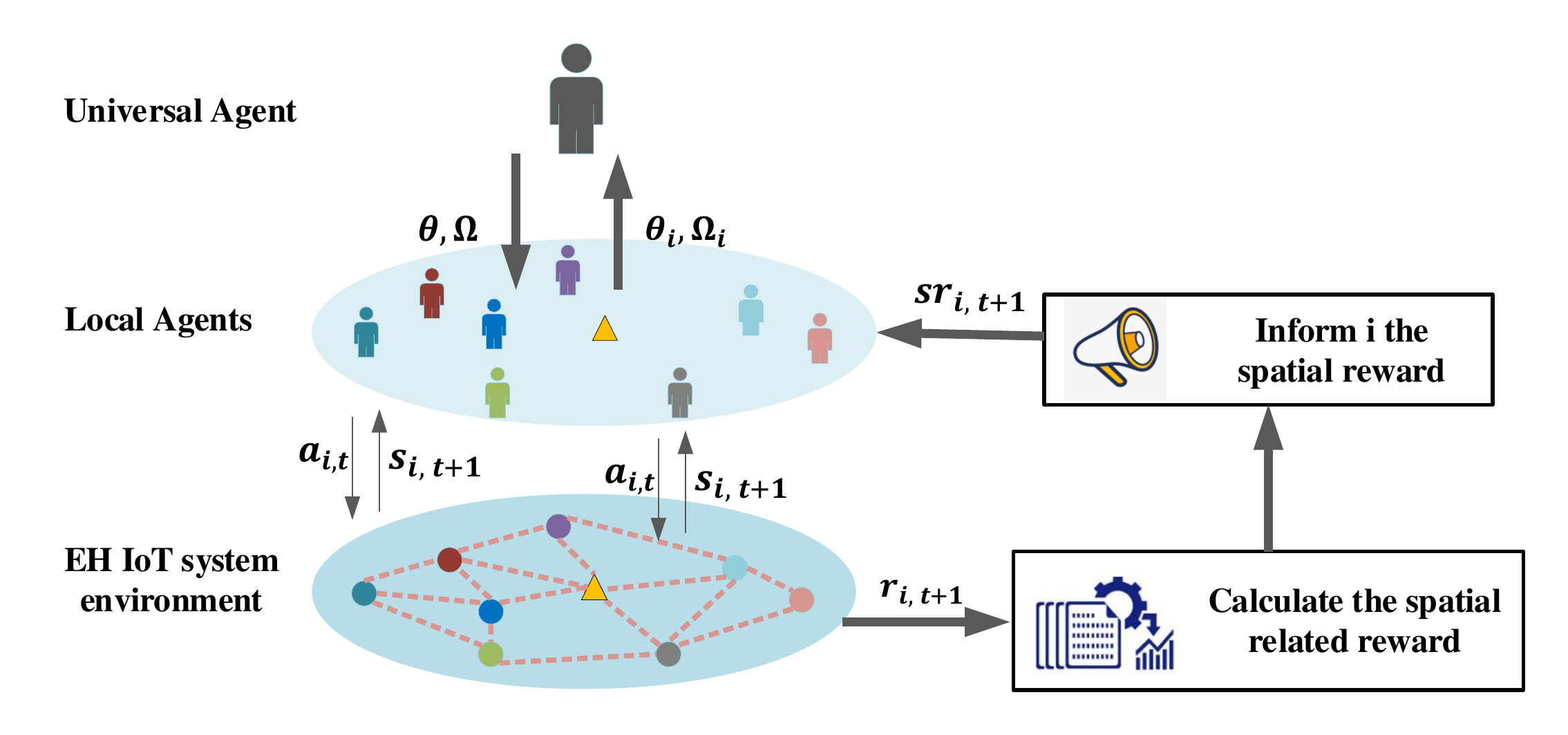}\vspace{-5pt}
\caption{Illustration of the proposed GAP algorithm.}
\label{Fig:2}\vspace{-15pt}
\end{figure}

To build a scalable and distributed algorithm, instead of using the single-agent DRL, we employ MARL to control each IoT device at local. The intuitive method is that each $i$ has a local agent which learns its own policy independently with the local experience information $(s_{i,t}, a_{i,t}, r_{i,t+1}, s_{i,t+1})$ at $t$. Each local agent concurrently interacts with the EH IoT system environment and asynchronously optimizes its local policy model $\theta_i$ as well as local V-function model $\Omega_i$(left side of~\figref{Fig:2}).
However, in multi-hop network the decision-making of local agents affect each others. To maximize the received data on Sink, local agents need to frequently exchange information and take cooperative actions which is unattainable due to the limited power budgets.
Although the local agent can not obtain the actions of other agents after deploying, it can obtain the rewards of other agents in the training period so that the agents can consider the influence of the actions of other agents. Therefore, instead of providing a local reward $r_{i,t+1}$ to the agent $i$, we design to feed a spatial global reward $sr_{i,t+1}$ to the agent $i$, which reflects the action influence to the whole system, as indicated in~\figref{Fig:2}.

To save computation resources, instead of training multiple local agents, we train an actual universal agent by creating virtual agents at local. Local agents concurrently and asynchronously optimize their local virtual models, $\theta_i$ and $\Omega_i$. After that, they update the universal agent.~\figref{Fig:2} shows the GAP overview, and the algorithm is described in Algorithm~\ref{agl:1}.

\begin{algorithm}[t]
\caption{GAP Training Procedure.}
\label{agl:1}
\footnotesize
\SetAlgoLined
\KwIn{$max\_episodes, \gamma, \iota_\theta, \iota_\Omega, n\_nodes, s\_time, e\_time, max\_step$}
 \KwOut{$\theta$, $\Omega$ }
    Initialize Actor network, $\theta$, with random weights; \\ 
    Initialize Critic network, $\Omega$, with random weights; \\ 
    Reset EH IoT system environment; \\
    $n\_episode \leftarrow 0$\\
\textbf{Create $n\_nodes$ threads} (Threads ID $i$ is from 0 to $n\_nodes-1$)\\
    \DontPrintSemicolon
    \SetKwBlock{DoParallel}{do in parallel}{end}
    \DoParallel{
      \While{n\_episode $<$ max\_episodes}{
        Download $\theta$, $\Omega$ to virtual local agent $\theta_i$, $\Omega_i$\\
        Reset local environment and obtain initial local state $s_{i,0}$\\
        $i\_step \leftarrow 0$;\\
         \For{$s\_time<real\_time<e\_time$}
        {Calculate the action $a_{i,t}$ by $\theta_i$\\
        Take action $a_{i,t}$ on the environment\\
        Obtain $r_{i,t+1}$ and $s_{i,t+1}$;Update $r_{i,t+1}$ to all gents;\\
        Calculate the spatial related reward $sr_{i,t+1}$\\
        Collect experience ($s_{i,t},a_{i,t},sr_{i,t+1},s_{i,t+1}$)\\
        $i\_step \leftarrow i\_step + 1$\\
         \If{$i\_step \geq max\_step$}{
        \textbf{Update} $\theta_i$ with $\iota_\theta \nabla \mathcal{L}(\theta_i)$~\eqref{eq:L_func_theta}  \\ 
        \textbf{Update} $\Omega_i$ with $\iota_\Omega \nabla \mathcal{L}(\Omega_i)$~\eqref{eq:L_func_omega}  \\ 
        \textbf{Update} $\theta_i$ and $\Omega_i$ to $\theta$ and $\Omega$, respectively\\
        }
        }
       
        $n\_episode \leftarrow n\_episode + 1$
    }}
\end{algorithm}  

\subsection{GAP Settings}
\label{sec:ia3cset}
In the multi-hop EH IoT system, each IoT devices has to decide on its energy allocation and next routing neighbor independently and simultaneously based on its local environment information. Such an EH IoT system scenario can be formulated as POMDP and we design GAP settings as follows:

\textbf{State} $s_i$: $s_i, i\in \mathcal{M}$ is the local environment state of agent $i$, which is denoted by $s_i = \{\textbf{nE}_i, \textbf{nQ}_i, nI_i\}$. $\textbf{nE}_i$ is the vector to describe the current energy of node $i$ ($E_{i,res}$) and its neighbors. $\textbf{nQ}_i$ represents the queue size of node $i$ and its neighbors. $nI_i$ is the source id of heading in $\textbf{queue}_i$, which is useful to prevent routing loop. 
  
\textbf{Action} $a_i$: The action consists of two members, defined as $a_i = \{\mathcal{E}_i, \mathcal{R}_i\}$. $\mathcal{E}_i \in [0,1]$ is to control energy allocation between sensing and transmission. If the current energy of device $i$($E_{i,res}$) is less than $\mathcal{E}_i*E_{i,max}$, the device $i$ stop its transmission and only perform sensing and energy harvesting. While $E_{i,res}>{E}_i*E_{i,max}$, it restart transmission operation. The second term $\mathcal{R}_i \in N_i$ aims to select relaying destination node for the current outgoing data packet on node $i$.
 
\textbf{Reward} $r_i$: The reward of $i$'s agent at time step $t$, $r_{i,t}$, is designed as below:
\begin{equation}\small\vspace{-5pt}
    r_{i,t} =\left\{
    \begin{aligned}
        & \Lambda + 1/k+Er,\text{~else~}\\
        & 0,\text{~if~} \mathcal{I}_i = i, h_i > H, \tau_{i,k}>T, E_{j,res}< 0, \sum_{l=1}^{k}A_{i,j}^l > \Delta
\label{eq:rw}
    \end{aligned}\vspace{-5pt}
    \right.
\end{equation}
where two cases of reward are described. In the first case, the data packet is transmitted successfully. $\Lambda$ is the one-step basic reward for agents to transmit the data packet out. $k$ is the number of the data packet being stored in $\textbf{queue}_j$, which encourages agents to send the task to the device that does not have a heavier transmission burden. The $Er$ is a bonus reward if the agent transmits data to Sink node, given as~\eqref{eq:Er}.
\begin{equation}
    Er =\left\{
    \begin{aligned}
        & log(total\ number\ of\ IoT\ devices*\Lambda),\text{~if~} j = \mathcal{O}\\
&0,\text{~else~}
\label{eq:Er}
    \end{aligned}
    \right.
\end{equation}

The second cases describe the device $i$ fails to transmit data. Five common failures are considered. (1) Routing loop: The data packet gets stuck in a loop and is send back to its source node $i$. (2) Hop expiration: The data packet with a limited budget $H$ on the number of routing hops. Without careful routing path design, the data packet is not delivered to Sink but gets stuck in relaying nodes that caused energy dissipation. (3) Time expiration: If the data packet stays in one node for a long time which might result in expired non-effectiveness data. (4) Receiver fails: Because of the transient nature of energy harvesting, the device at the destination might be offline. 
(5) Queue overflow: The storage of destination node $j$ reaches its capacity $\Psi$. The packet will be refused by the destination but resent by the source node.

Eq.~\eqref{eq:rw} encouraged Sink neighbors to transmit data to Sink obtaining a large reward. But for those nodes that are not Sink neighbors, they are only encouraged to transmit data to the device have low transmission burden($Er=0$). It is unclear if the transmission is useful for data delivery to Sink. Thus, a global spatial related reward $sr_{i,t}$ is designed as $sr_{i,t} = r_{i,t}+ \sum_{\tau'=t-1}^{t}~\sum_{j\neq i,\forall j\in\mathcal{M}}(1/d_{i,j}*r_{j,\tau'})$, where $d_{i,j}$ is the distance between $i$ and $j$. $sr_{i,t}$ represents an spatial reward accumulation of all agents from time $a_{i,t-1}$ to time $a_{i,t}$.

\section{Evaluation}
\label{sec:4}
\begin{figure}[t]
\vspace{-10pt}
\centering
\includegraphics[width=0.45\textwidth]{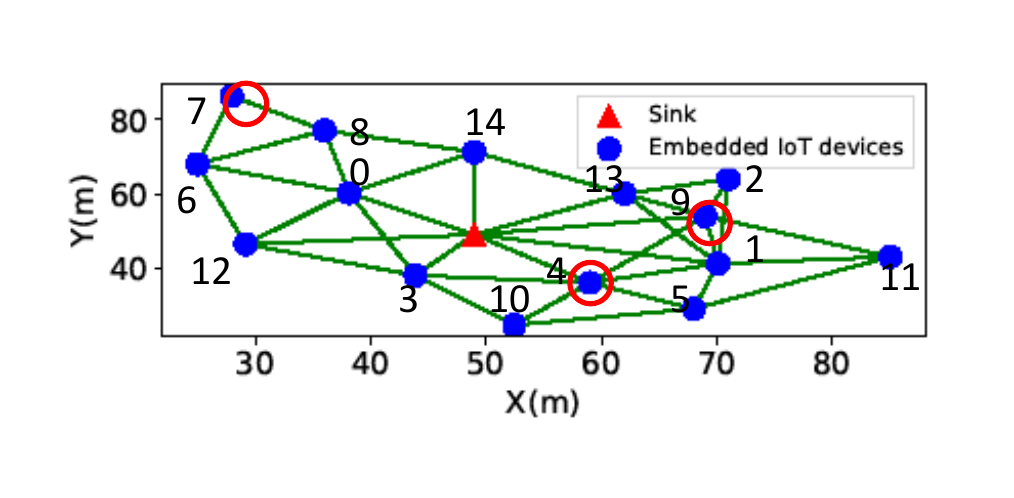}\vspace{-20pt}
\caption{Topology of 15-node EH IoT network.}
\label{Fig:3}
\vspace{-15pt}
\end{figure}

We developed a solar energy powered~\cite{database} IoT network simulator to evaluate our proposed GAP algorithm. \figref{Fig:3} shows the topology of a 15-node example of multi-hop EH IoT system. The network consists of a single Sink (data transmission destination) and multiple EH IoT nodes.

\subsection{Experimental Settings}
\textbf{EH IoT network.}
We simulate our multi-hop EH IoT network as~\figref{Fig:1} in 50 days. Each day is running from 8:00 to 17:00. For fairness comparison, at the beginning of each day, the environment will be reset. To ensure the sufficient start-up energy, $E_{i,res}$ is initialized to energy storage capacity ($1J$).~\tabref{table1} lists the parameters used in our simulation. 

\textbf{GAP algorithm.} 
We train GAP for $50$ episodes, which matches $50$ days in our EH IoT network simulation. We set discount factor $\gamma = 0.9$ and the learning rate $\iota_\theta = 1e-4$ and $\iota_\Omega = 3e-4$ for Actor $\theta$ and Critic $\Omega$, respectively. 
The architecture of the Actor (policy) neural network model and Critic (value) neural network model are 49-256-512-256-128 and 49-512-1024-256-1. Note that the single output of Critic is the estimated Q-value.
Agent can select the energy threshold $\mathcal{E}_i$ from $[0, 0.3, 0.6, 0.9]$. In case the time interval is too long while optimizing the neural network after each day ending, the neural network model is optimized for every $50$ steps. 

\textbf{Baseline.} 
We compare our GAP algorithm with three baselines.
\textbf{1) GAPR} Compared with GAP, GAPR makes decision on routing selection only; 
\textbf{2) Q-table} baseline will deploy a Q-table in each EH IoT devices. Since one centralized Q-table is impractical, we did not adopt it; 
\textbf{3) ESDSRAA} baseline allocates energy budget and decides a constant sensing rate per hour. After energy allocating, it decides the relaying node based on the delivery rate and geo-information of neighbors.

\textbf{Measurement.}
We evaluate the algorithms performance from two perspectives: \textbf{efficiency} and \textbf{effectiveness}. We adopt the \textit{total reward}, the \textit{Sink received data} and the \textit{delivery rate} across all agents w.r.t. training episodes as our measurements on efficiency. Specifically, the delivery rate is defined as $\frac{\textit{Sink received data size}}{\textit{the total sensed data size}}$. The reward is measured as a natural scoring mechanism. We listed detail on the amount of sensed data, received data and transmitted data of several nodes and their corresponding energy cost to analyze effectiveness.

\begin{table}[t] 
\centering\small
\caption{EH IoT Network Parameters}
\vspace{-5pt}
\begin{tabular}{p{0.84cm}p{4cm}p{2cm} } 
  \hline
\textbf{Notation} & \textbf{Definition}  & \textbf{Value/Range} \\ 
  \hline
$\zeta$ & transmission range  & $25$m  \\ 
  $A_{i,j}$ & size of data packet & $[3720,5120]$bits\\
  $Pr$ & process speed & $80 $bits/s\\
    $\Psi$ & maximum queue size& $15*5120$bits\\
  $P_{i,trans}$ & transmission power & $0.1$w  \\
   $P_{i,recei}$ & receiving power  & $0.05$w \\
  $P_{i,sleep}$ & sleeping power & $0.0005$w\\
     $P_{i,sense}$ & sensing power & $0.01$w\\
  $E_{i,max}$ & energy storage capacity & $1$J\\
   $\Lambda$ & one-step basic reward & $0.1$\\
   T & expiration time of data packet & $1800$s\\
   H & maximum experienced hops & $8$\\
   \hline 
\small
\end{tabular}
\label{table1}\vspace{-20pt}
\end{table}

\subsection{Analysis and Discussion}

\subsubsection{\textbf{Efficiency}}
\begin{figure*}[t]
\centering
\begin{minipage}[t]{0.3\textwidth}
\centering
\includegraphics[width=5.5cm]{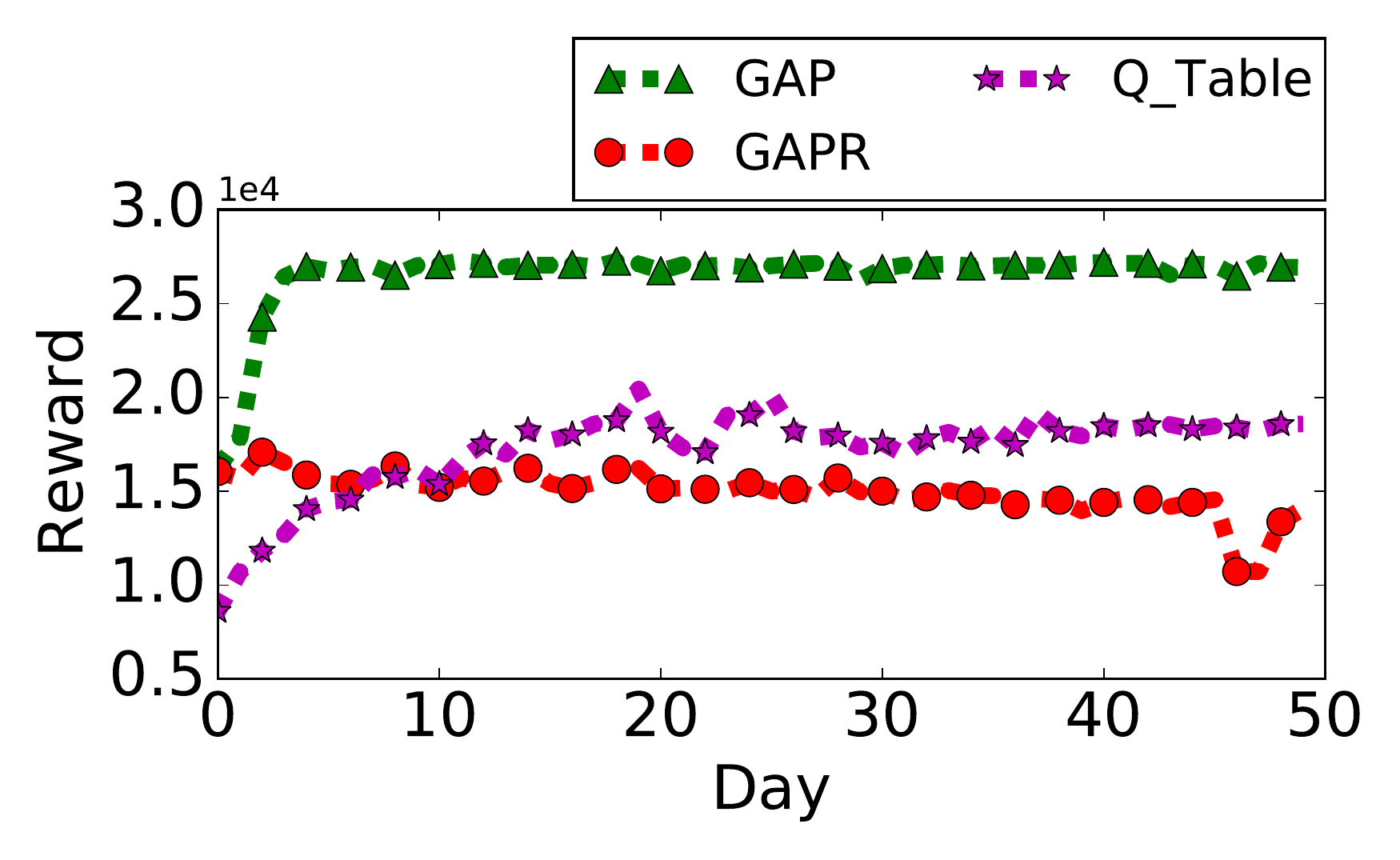}
\vspace{-20pt}
\caption{Total obtained reward each day.}
\label{Fig:Rw}
\end{minipage}
\begin{minipage}[t]{0.3\textwidth}
\centering
\includegraphics[width=5.5cm]{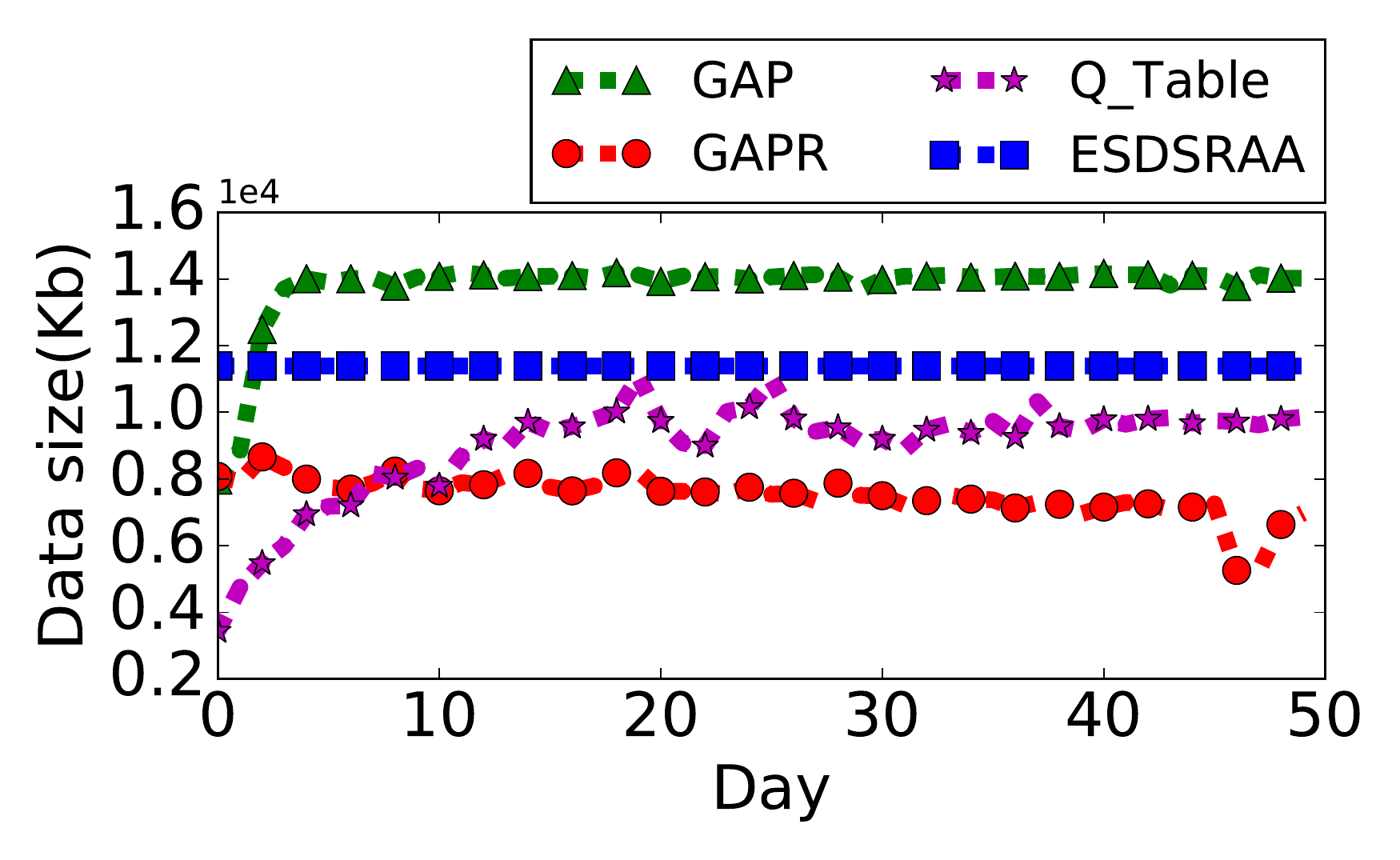}
\vspace{-20pt}
\caption{Sink received data each day.}
\label{Fig:Sk}
\end{minipage}
\vspace{-10pt}
\begin{minipage}[t]{0.3\textwidth}
\centering
\includegraphics[width=5.5cm]{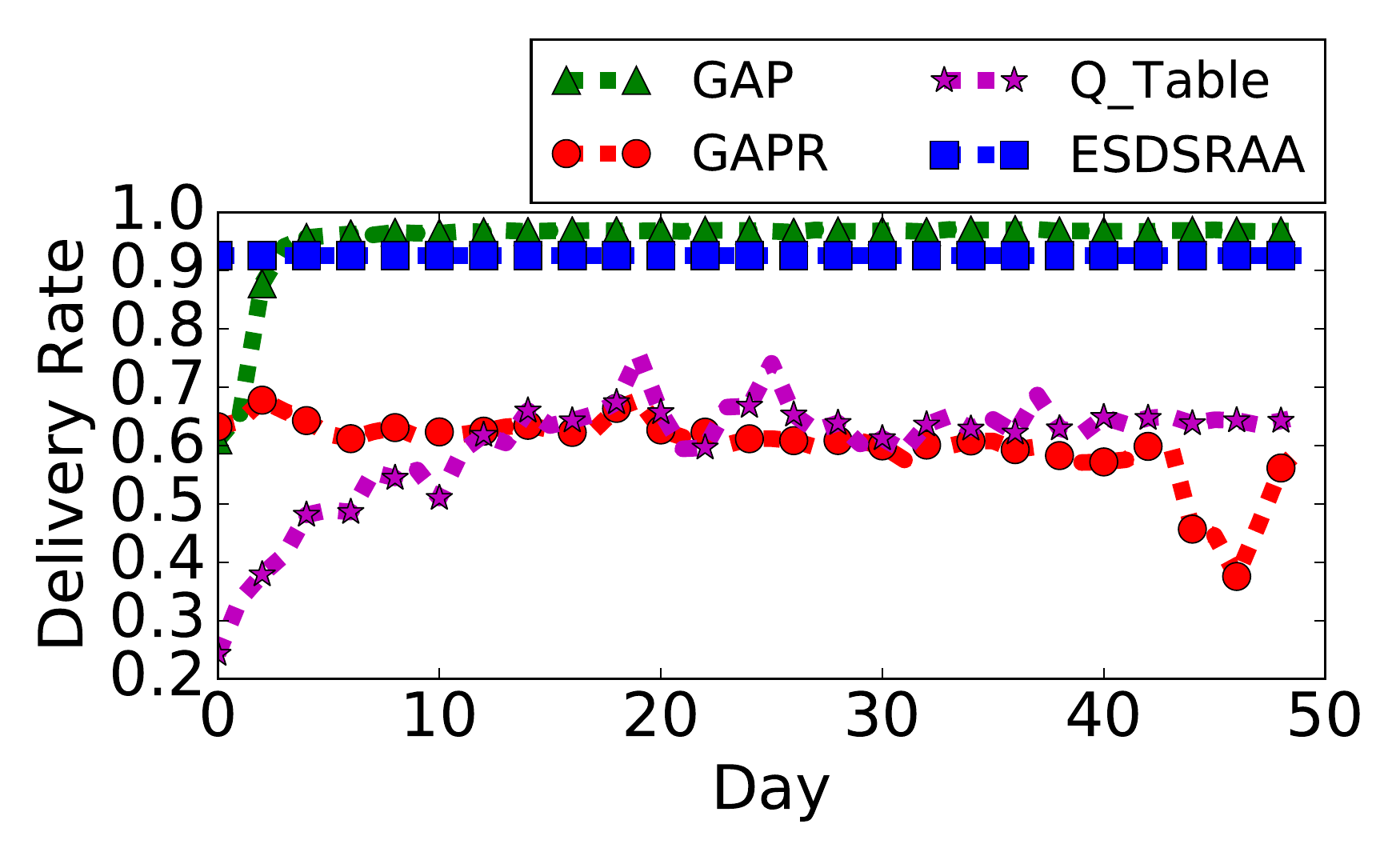}
\vspace{-20pt}
\caption{Delivery rate each day.}
\label{Fig:Rate}
\end{minipage}
\end{figure*}

As shown in~\figref{Fig:Rw}, our GAP agent firmly and increasingly earns rewards in the first 5 training days. It surpasses the Q-table and GAPR significantly. During the training period, the neural network is optimized every 50 steps. Although it converges fast within 5 days, the neural network model goes through a large amount of data each day. Since ESDSRAA is the traditional method to calculate the energy allocation and routing selection strategy, it dose not have a reward parameter. The Q-table is convergent at $10^{th}$-day and obtained half reward of GAP agent. This result indicates the best learning efficiency of the proposed GAP. To evaluate the agent's efficiency on data delivery, we further measure the total Sink node received data on each training day in~\figref{Fig:Sk}. GAP agent achieved $14.18$Mb data, where GAPR, Q-table, and ESDSRAA is at $7.29$Mb, $11.10$Mb, $11.39$Mb, respectively. The proposed GAP improved 24.50\% Sink node received data compared with the best baseline ESRSDAA.

If we only evaluate the network performance based on the received data on Sink, the EH IoT devices might ``play a tricky''. They can generate a huge number of data but routing a few portions of generated data, which is not a fair comparison. Thus, to scientifically evaluate the performance, we measure the delivery rate.~\figref{Fig:Rate} indicated the delivery rate achieved by GAP (0.95) and ESDSRAA (0.93). Although they have a similar delivery rate, in~\figref{Fig:Sk}, GAP has a higher Sink received data. After jointly considering the delivery rate and Sink received data amount, the result proves the proposed GAP agent can generate an outstanding routing policy. Q-table and GAPR complete almost half Sink received data and half rate of GAP. We will figure out the reason in~\secref{sec:eff}. Moreover, the curve trend of reward in~\figref{Fig:Rw} is very similar with~\figref{Fig:Sk} and~\figref{Fig:Rate}, which further proves the reward design is suitable in our case.

\begin{figure*}[t]
\centering
\begin{minipage}[t]{0.3\textwidth}
\centering
\includegraphics[width=5.5cm]{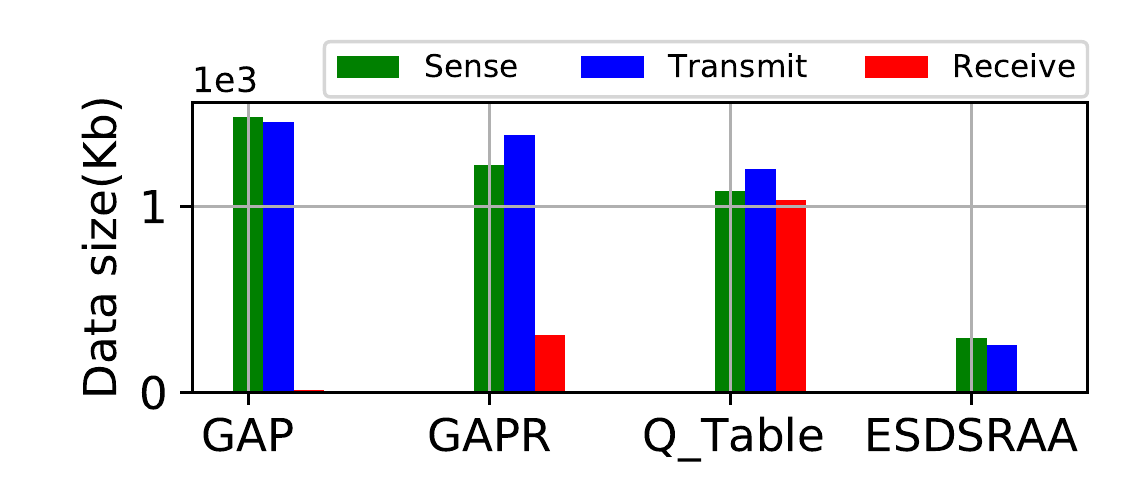}
\vspace{-20pt}
\caption{Data detail on node-7.}
\label{Fig:Data1}
\end{minipage}
\begin{minipage}[t]{0.3\textwidth}
\centering
\includegraphics[width=5.5cm]{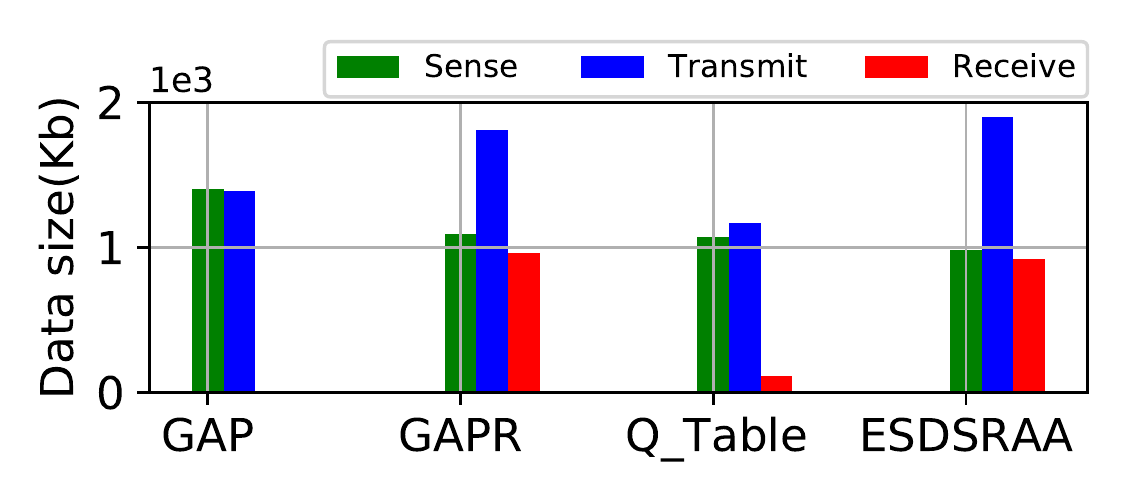}
\vspace{-20pt}
\caption{Data detail on node-4.}
\label{Fig:Data2}
\end{minipage}
\vspace{-10pt}
\begin{minipage}[t]{0.3\textwidth}
\centering
\includegraphics[width=5.5cm]{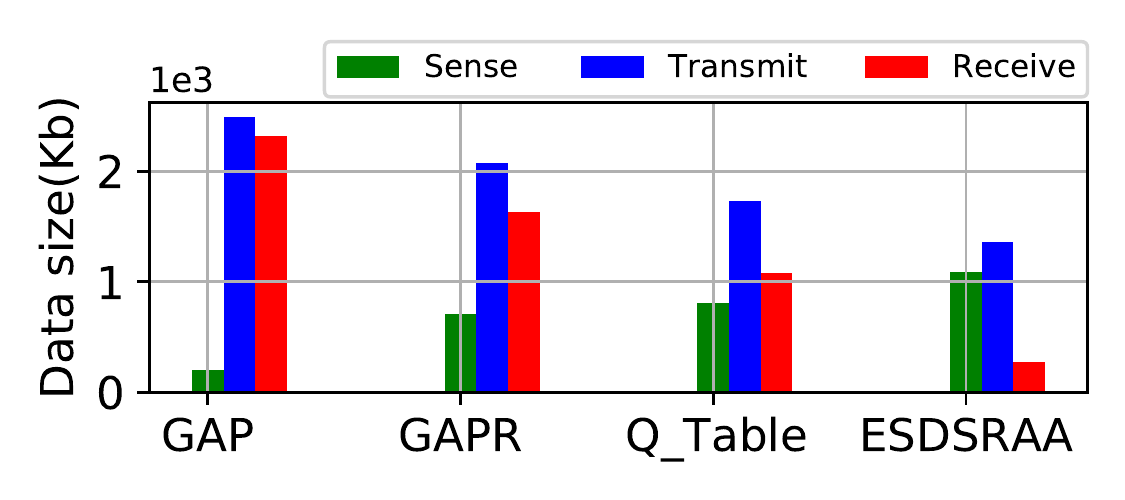}
\vspace{-20pt}
\caption{Data detail on node-9.}
\label{Fig:Data3}
\end{minipage}
\end{figure*}

\begin{figure*}[t]
\centering
\begin{minipage}[t]{0.3\textwidth}
\centering
\includegraphics[width=5.5cm]{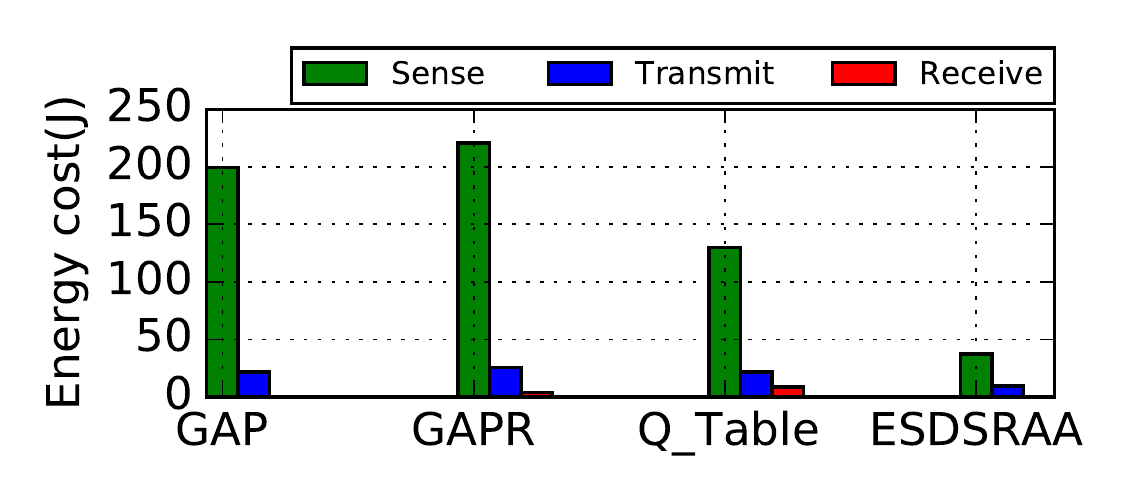}
\vspace{-20pt}
\caption{Energy cost on node-7.}
\label{Fig:Energy1}
\end{minipage}
\begin{minipage}[t]{0.3\textwidth}
\centering
\includegraphics[width=5.5cm]{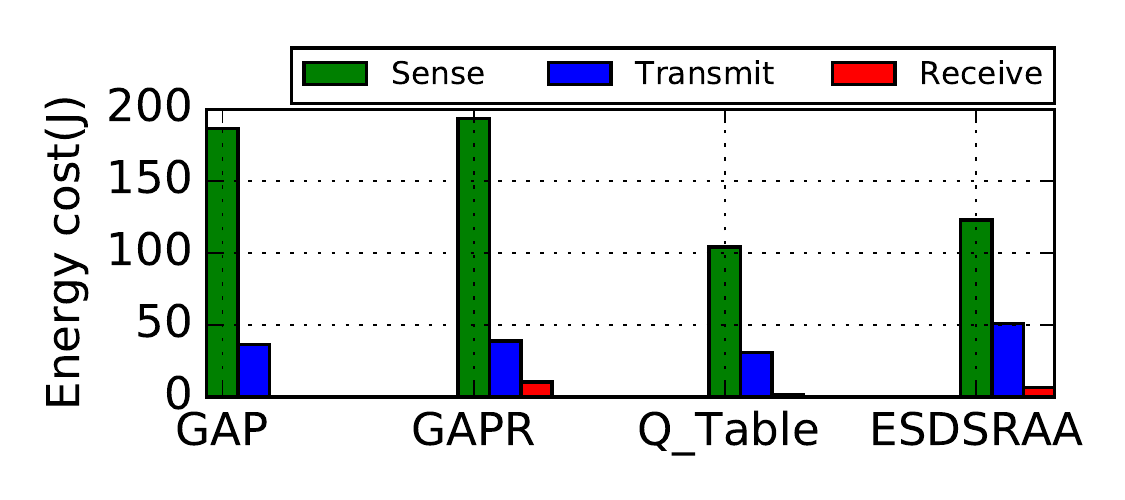}
\vspace{-20pt}
\caption{Energy cost on node-4.}
\label{Fig:Energy2}
\end{minipage}
\vspace{-10pt}
\begin{minipage}[t]{0.3\textwidth}
\centering
\includegraphics[width=5.5cm]{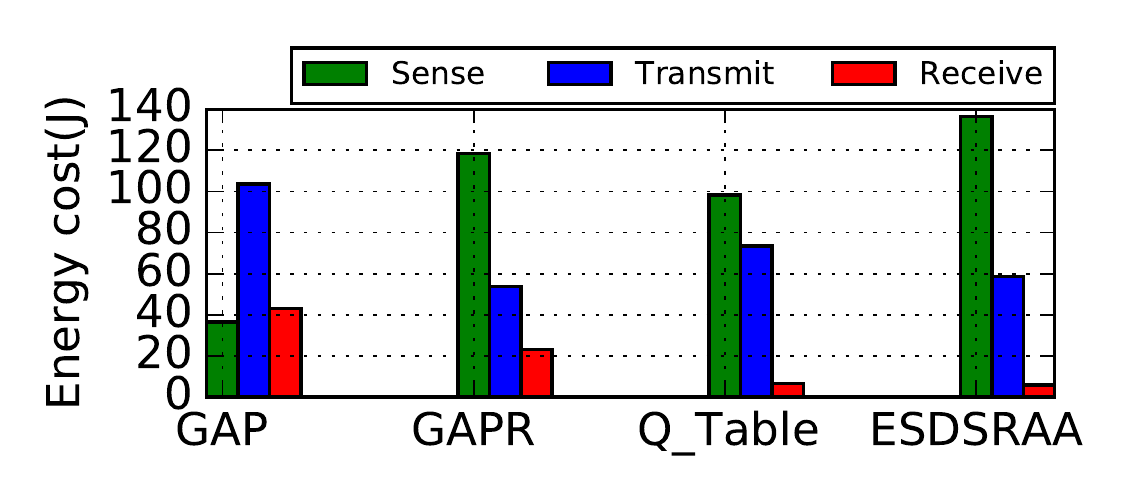}
\vspace{-20pt}
\caption{Energy cost on node-9.}
\label{Fig:Energy3}
\end{minipage}
\end{figure*}

\subsubsection{\textbf{Effectiveness}}\label{sec:eff}

From~\figref{Fig:Data1} to~\figref{Fig:Energy3} we listed the data size and their energy cost details on sensing, transmission, and receiving on the $50^{th}$ training day for three nodes. Without loss of generality, we select node-7, node-4, and node-9, labeled in~\figref{Fig:3} that presents three major kinds of nodes. Node-7 is located at the corner of the network, which rarely relays for others; Node-4 is located at the network center, which can directly communicate with Sink; Node-9 is a neighbor of node-4, which affects the decision-making of node-4.

Intuitively, due to its position, node-7 should not receive data. As shown in~\figref{Fig:Data1}, GAP and ESDSRAA agents did not receive any data on node-7. However, for GAPR and Q-table, node-7 has an unanticipated receiving data, which means there is a routing loop or agent complete a wrong routing selection resulting in energy dissipation. Therefore it implies Q-table and GAPR should complete fewer data delivery tasks. It is exactly proved by~\figref{Fig:Sk} and ~\figref{Fig:Rate}, where both Sink received data and delivery rate is half that of GAP.

In~\figref{Fig:Data2}, data size of sensing, transmitting, and receiving on node-4 are similar to node-7. However, node-4 is a central node that is close to Sink (one-hop). While the node located near Sink should have a heavier burden on relaying, it should relay data for the edge nodes. But node-4 did not receive data in GAP algorithm. For nodes near Sink such as node-4, even that they sense plenty of data, the sensed data can not be totally delivered to Sink. Because it has a limited battery capacity, if it still costs much energy on sense, there is insufficient energy for transmission. However,~\figref{Fig:Sk} indicated GAP delivered the most data to Sink. The reason is that relaying task is completed by other nodes which are also near Sink. To further validate it, we also listed the details on node-9 in~\figref{Fig:Data3}. It indicated the transmission data is dramatically higher than the sensing data on node-9 for GAP agent.
While nodes have heavier transmission/relaying task that is related to neighbor numbers and nodes positions, sensing operation is anticipated to be reduced so that nodes can save more energy for transmission. 

If the relaying missions are not heavy, the node should sense data as much as possible for sensing quality(node-9 of GAP). However, the overall distribution on sensing, receiving and transmitting of GAPR, Q-table, ESDSRAA in~\figref{Fig:Data1}-~\figref{Fig:Data3} is generally similar. But, the ratio of sensing and transmission for GAP is varied with different nodes. The result of GAP in~\figref{Fig:Data1}-~\figref{Fig:Data3} implies its ability to dynamically adapt the sensing and transmission ratio for the system's global optimal goal with the consideration of the nodes local information.

Received data pluses sensed data should be equal to transmitted data. Otherwise, the packet drop occurs. In~\figref{Fig:Data1}-~\figref{Fig:Data3}, this is completed by GAP and ESDSRAA. GAPR and Q-table did not achieve it because of packet drop caused by routing loop or packet expiration, which implies wrong routing selection. GAPR only makes decisions on routing. Better routing selection is expected. However, its routing performance is depleted. It further illustrates jointly considering energy allocation and routing selection is critical.

In~\figref{Fig:Data1}, the transmission data size of GAPR is less than the transmission data size of GAP. However, in~\figref{Fig:Energy1} the energy consumption for transmission of GAPR is more than that of GAP, which means re-transmission happened. That is another manner that caused the energy dissipation. A similar situation also happens in~\figref{Fig:Energy3} while we compare transmission energy cost of Q-table and GAPR. Moreover, the total energy cost in the other three algorithms is much more than the energy cost in GAP on each node, which means they have more opportunities for harvesting energy. More Sink received data should be achieved. But~\figref{Fig:Sk} reflects they did not.

\section{Conclusion}
\label{sec:conclusion}

This paper constructs a comprehensive multi-hop data transmission model for EH distributed IoT networks. With the objective of maximizing the amount of Sink received data, we proposed GAP establishing a distributed MARL for EH IoT networks to jointly optimize the routing policy and energy allocation with the consideration of global long-term optimality and devices entanglement. The spatial global reward is integrated to MARL to address the entangled problem in the multi-hop EH IoT system. The experimental results indicated the outstanding performance of the proposed GAP compared with the three baselines. 

\vspace{-5pt}
	\bibliographystyle{IEEEtran}
	\bibliography{IEEEabrv,main10272020}


\end{document}